# Successive Symmetry Breaking in a $J_{\text{eff}} = 3/2$ Quartet in the Spin–Orbit Coupled Insulator $Ba_2MgReO_6$

Daigorou Hirai[1]* and Zenji Hiroi[1]

[1] *Institute for Solid State Physics, University of Tokyo, Kashiwa, Chiba 277-8581, Japan*
*e-mail address: dhirai@issp.u-tokyo.ac.jp*

We report on the cubic double perovskite $Ba_2MgReO_6$ containing $Re^{6+}$ ions with the $5d^1$ electron configuration. Resistivity, magnetization, and heat capacity measurements using single crystals show that the compound is a Mott insulator with a magnetic transition at $T_m = 18$ K, which is accompanied by a weak ferromagnetic moment with [110] anisotropy. Another transition is observed at $T_q = 33$ K in heat capacity, where the inverse of magnetic susceptibility changes its slope, indicating a substantial change in the electronic state. The significance of spin–orbit coupling is revealed by the reduced effective magnetic moment of ~$0.68\mu_B$ at high temperature above $T_q$ and the total electronic entropy close to $R\ln 4$. These features indicate that $Ba_2MgReO_6$ is a spin–orbit coupled Mott insulator possessing a $J_{\text{eff}} = 3/2$ quartet state, which exhibits quadrupolar and dipolar orders at $T_q$ and $T_m$, respectively.

## I. INTRODUCTION

Materials with strong spin–orbit coupling (SOC) and large electron correlation have received great interests in recent years [1], because they can exhibit a variety of exotic phases such as spin liquids [2–4], Weyl semimetals [5–7], axion insulators [8,9], topological Mott insulators, and multipolar orders [1,10,11]. Particularly, in the limit of strong electron correlation, the spin–orbit coupled Mott insulator appears. The first example was reported in the iridium oxide $Sr_2IrO_4$ [12,13]. In this compound, a spin–orbit coupled $J_{\text{eff}} = 1/2$ state is half-occupied and splits into two by electron correlations with the lower state fully occupied to induce a spin–orbit-entangled Mott insulating state [12].

In order to explore novel quantum phases stabilized by SOC, materials search has been carried out especially for $5d$ transition metal compounds. In $5d$ transition metal compounds, the spin $S$ is no longer a good quantum number and coupled with the orbital angular momentum $L$ via SOC, and, thus, the ground state becomes a multipolar state represented by the total angular momentum $J$. For example, in the case of $5d^5$ ion placed in a cubic crystal field, as seen in iridates, the $t_{2g}$ manifold of the $d$ electron is triply degenerate and carries an effective angular momentum of $l_{\text{eff}} = 1$. This degeneracy is lifted by SOC into two states with $J_{\text{eff}} = l \pm s = 3/2$ and $1/2$. The $J_{\text{eff}} = 1/2$ ($\Gamma_7$) doublet is chosen as the ground state, and the $J_{\text{eff}} = 3/2$ ($\Gamma_8$) quartet appears at higher energy when the crystal field splitting and SOC compete with each other [14]. On the other hand, the excited and ground states are reversed for the $5d^1$ ion, resulting in the $J_{\text{eff}} = 3/2$ quartet as the ground state.

In $Sr_2IrO_4$, which comprises octahedrally coordinated $Ir^{4+}$ ions in the $5d^5$ electron configuration, a $J_{\text{eff}} = 1/2$ ($\Gamma_7$) state has actually been evidenced by resonant X-ray scattering experiments. [13] For other iridates with $Ir^{4+}$ ions such as $Na_2IrO_3$ [15], a new type of spin liquid called the Kitaev spin liquid is predicted, in which highly anisotropic interactions between the $J_{\text{eff}} = 1/2$ pseudo-spins, far different from conventional isotropic Heisenberg interactions between simple spins, occur in the honeycomb lattice of $Ir^{4+}$ ions [2,4]. On the other hand, in the case of single $d$ electron placed in an octahedral crystal filed, it is theoretically expected that a $J_{\text{eff}} = 3/2$ ($\Gamma_8$) quartet state causes complex multipolar orders arising from the large degeneracy [1,10] To materialize this intriguing physics for the $5d^1$ system, some double perovskite (DP) compounds have been studied thus far. Here we focus one of them, $Ba_2MgReO_6$ with $Re^{6+}$ ions in the $5d^1$ electron configuration.

Elpasolite ($K_2NaAlF_6$) type compounds [16] provide an ideal playground for exploring the above-mentioned novel quantum phases originating from SOC and electronic correlation. They crystallize in an ordered double perovskite (DP) structure with the general formula $A_2BB'O_6$, where the B site in the perovskite $ABO_3$ is occupied by two kinds of cations in the rock salt manner. We focus the DP compounds having non-magnetic B cations and $4d$ or $5d$ transition metals at the B′ site. They have two advantageous characteristics: one is the Mott insulating state, and the other is the high symmetry of the B′ site. Most of $5d$ transition metal compounds are weakly correlated metals because of relatively large band widths coming from extended $5d$ orbitals. For instance, $ReO_3$, which is composed of corner-sharing $ReO_6$ octahedra, is one of the highly conducting oxides. [17] In contrast, DPs tend to be Mott insulators because of the small electron transfer over the long interatomic distances between transition metal ions, which may be smaller than electron correlations. This structural feature provides a suitable playground for studying the physics of correlated electrons. Concerning the second advantage of the high symmetry at the transition metal site, one can study the purely-electronic instability of the $J_{\text{eff}} = 3/2$ quartet in the absence of built-in structural distortion in DPs having a cubic structure of the space group $Fm\text{–}3m$.

For the DPs with the $d^1$ electronic configuration, a novel multipolar order is theoretically predicted [10]. Derived from the $J_{\text{eff}} = 3/2$ quartet are dipole, quadrupole, and octupole moments. According to the theoretical work using a mean-field approach by Chen, Pereira, and Balents, two transitions occur successively upon cooling: first, a non-magnetic quadrupolar order and second, a magnetic order with a ferromagnetic moment; the quadrupolar order is stabilized by thermal fluctuations at a higher temperature. The magnetic order is composed of two sublattices each with ferromagnetically aligned moments lying in the (001) plane [18], as depicted in Fig. 1. The direction of the magnetic moments rotates by an angle of $+\phi$ in one sheet and by $-\phi$ in the next sheet with respect to the [110] direction; thus, an uncompensated moment appears along [110]. This spin arrangement has been called the canted ferromagnetic order [18,19], but, in the present study, we call it the canted antiferromagnetic (CAF) order following the convention as it consists of two sublattices.

This type of unusual magnetic order has actually been observed below 6.3 K in the cubic DP $Ba_2NaOsO_6$ with the $Os^{7+}$ ion in the $5d^1$ electronic configuration [19]; however, the presence of a quadrupolar order at high temperatures in the compound remains controversial, as will be mentioned in the next paragraph. The magnetic order is characterized by a small saturation moment of $\sim 0.2\mu_B$ (the Bohr magneton) [20,21] and easy-axis anisotropy along the [110] direction [21], which is difficult to be explained by the Landau theory for a cubic ferromagnet. Recent nuclear magnetic resonance (NMR) study by Lu et al. revealed that the order is not a simple ferromagnetic order but in fact a CAF order with a large spin canting angle of $\phi = 67°$ [19].

Ordering of a quadrupolar degree of freedom is generally more difficult to detect than dipole order. However, because a quadrupolar moment, which is an anisotropic distribution of electronic charge, can couple with the lattice and thus induce a lattice distortion, there is a possibility to obtain indirect evidence by structural characterization. For $Ba_2NaOsO_6$, a transition from a high-temperature cubic to a low-temperature tetragonal structure was reported at 320 K by means of high-resolution synchrotron XRD and was claimed to be a quadrupolar order [18,19], though no experimental data was given. A structural distortion was actually detected by the NMR study as a splitting of the spectra below 12 K, much lower than 320 K, just above the magnetic order [19]. It was suggested that the lattice distortion accompanied by the quadrupolar order is too small to detect at high temperatures and becomes detectable by NMR within the resolution of 0.01%.

The $Re^{6+}$ ion also hosts the $5d^1$ electronic configuration, as is the case for $Os^{7+}$; the $W^{5+}$ ion can be another example, but there are few DP compounds. In the DP, the B′ site can adopt $Re^{6+}$ ions when both of the A and B sites are occupied by divalent cations; $A_2BReO_6$ (A = Ba, Sr, Ca; B = Mg, Ca, Sr, Ba, Zn, Cd) [16,22–30]. Among them, the compounds having $Ba^{2+}$ at the A site, $Ba_2BReO_6$ (B = Mg, Ca, Zn, Cd), crystallize in the cubic (space group $Fm\text{–}3m$) structure at room temperature, while the others with smaller A cations crystallize in distorted structures, as is often observed for the normal perovskites. Structural studies have been carried out for all the Ba compounds except for $Ba_2CdReO_6$: $Ba_2CaReO_6$ shows a transition to a tetragonal structure (space group $I4/m$) at 120 K [29], while $Ba_2MgReO_6$ and $Ba_2ZnReO_6$ likely remain cubic down to 3.5 K [30].

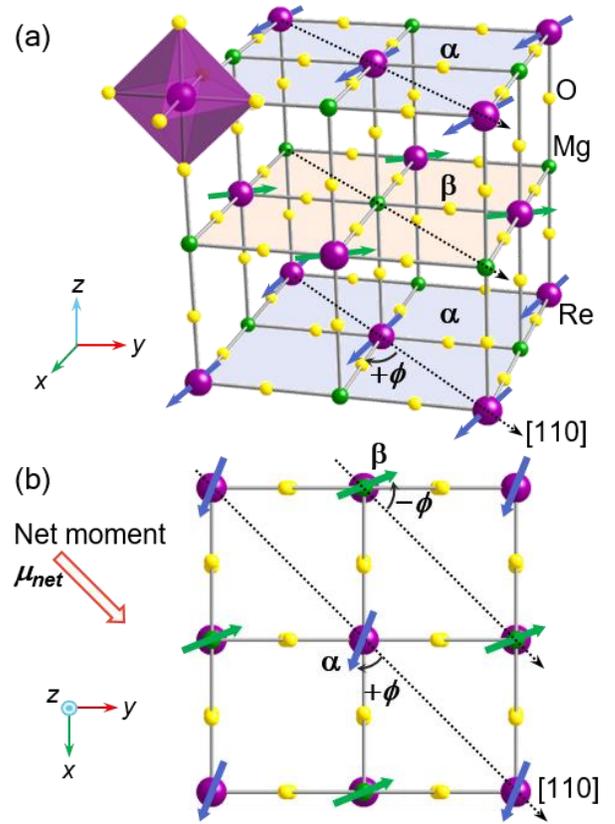

Fig. 1. (Color online) (a) Perspective view of the crystal structure of the double perovskite $Ba_2MgReO_6$ [space group $Fm\text{–}3m$, $a = 8.085(5)$ Å] and (b) its projection along the [001] direction. Mg (small green sphere) and Re (large purple sphere) atoms have the rock salt type order with oxygen atoms (small yellow sphere) between them. Ba atoms at the centers of the octant cubes are not shown for clarity. Blue and green arrows at the Re atoms represent the spin orientations in the theoretically proposed spin model in ref. 18, which is experimentally established by NMR for $Ba_2NaOsO_6$ [19] and is probably realized in $Ba_2MgReO_6$.

We focus on $Ba_2MgReO_6$ as a promising candidate for realizing a quadrupolar order in the spin–orbit coupled insulator with $5d^1$ electron configuration. The synthesis, crystal structure, and magnetic properties of $Ba_2MgReO_6$ were reported by Bramnik et al. in 2003 for the first time [28]. In 2016, detailed characterizations of the physical properties using polycrystalline samples (XRD, neutron diffraction, magnetic susceptibility, heat capacity, muon spin rotation, inelastic neutron scattering measurements) were done by Marjerrison et al. [30]. According to these studies, $Ba_2MgReO_6$ is a ferromagnet with $T_C$ = 18 K with a small saturation moment of $0.3\mu_B$ at 3 K, which resembles the magnetic properties of $Ba_2NaOsO_6$. A notable observation in the previous heat capacity measurements is the broad peak at around 33 K in addition to the sharp peak at the magnetic transition. The origin is not clear but may not be a structural transition as their neutron diffraction measurements did not detect a lattice distortion. It is also noted that there is no corresponding anomaly in magnetic susceptibility at 33 K. In these previous studies, crucial information on anisotropy in properties is missing because of the lack of single crystals.

Here, we study the physical properties of single crystalline $Ba_2MgReO_6$ by resistivity, magnetization, and heat capacity measurements. A Mott insulating state with an activation energy of 0.17 eV is revealed, and a magnetic order with a ferromagnetic moment below $T_m$ = 18 K is confirmed. Our anisotropy measurements reveal a CAF order with a net moment of 0.25 ~ $0.30\mu_B$ along the [110] direction, which is consistent with the theoretical prediction [18] and is similar to that of $Ba_2NaOsO_6$. The $J_{eff}$ = 3/2 state of $Ba_2MgReO_6$ is evidenced by the highly reduced effective magnetic moment of $0.7\mu_B$ and the total magnetic entropy of 11.3 J K$^{-1}$ mol$^{-1}$ close to $R$ln4. A clear anomaly is observed in the heat capacity at $T_q$ = 33 K indicating a bulk phase transition. Interestingly, the inverse of magnetic susceptibility changes its slope at $T_q$, which implies a substantial change in the magnetic interactions. Combined with the theoretical prediction, our finding demonstrates that the transition at $T_q$ is likely a quadrupolar transition.

## II. EXPERIMENTAL

Black shiny single crystals of $Ba_2MgReO_6$ were grown by the flux method. A stoichiometric mixture of BaO, MgO, and $ReO_3$ powders was mixed with a flux composed of 36 wt% $BaCl_2$ and 64 wt% $MgCl_2$ in an argon-filled glove box. The mixture was sealed in a platinum tube of 20 mm long and 6 mm in diameter. The tube was heated at 1300 °C and then slowly cooled to 900 °C at a rate of 5 °C/h, followed by furnace cooling to room temperature. After the residual flux was washed away by distilled water, several crystals were obtained, which have the truncated octahedral morphology with the maximum size of 3 mm, as typically shown in Fig. 2.

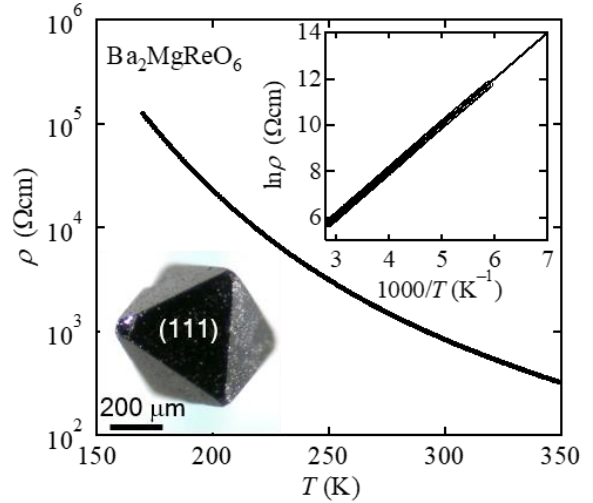

Fig. 2. (Color online) Temperature dependence of resistivity $\rho(T)$ of a $Ba_2MgReO_6$ single crystal. The upper inset shows the Arrhenius plot: log$\rho(T)$ vs. $T^{-1}$. The lower inset shows a photograph of a typical single crystal of $Ba_2MgReO_6$ having the triangle (111) facet.

The chemical composition of single crystalline samples was confirmed to be close to the stoichiometry within the experimental error by energy dispersive X-ray spectroscopy in a scanning electron microscope (SEM-EDX, JEOL JSM-5600). The crystal structure was characterized by means of x-ray diffraction conducted in a R-AXIS RAPID IP diffractometer with a monochromated Mo K$\alpha$ radiation at 300 K. The X-ray diffraction pattern was indexed based on an F-centered cubic cell with a lattice parameter of $a$ = 8.085(5) Å, which is consistent with the previous reports on powder samples [28,30].

The magnetic susceptibility was measured in a magnetic properties measurement system (MPMS-III, Quantum Design). A crystal of 3 mm × 2.5 mm × 2 mm and weight of 45 mg attached on a quartz sample holder by varnish was used for the measurements. The electrical resistivity was measured by the standard four-probe method with silver paste (DuPont 4922) as the electrical contact in a physical properties measurement system (PPMS, Quantum Design). Heat capacity measurements were performed by the relaxation method in the PPMS.

## III. RESULTS

### 1 Electrical resistivity

Electric resistivity measurements clearly reveal the insulating nature of $Ba_2MgReO_6$. As shown in Fig. 2, the resistivity is large, approximately 1 MΩ cm, at room temperature and increases exponentially with decreasing temperature, as evidenced by the linear behavior in the Arrhenius plot [log$\rho(T)$ vs. $T^{-1}$] in the inset of Fig. 2. The deduced thermal activation energy is $\Delta$ = 0.17 eV, which is comparable to 0.26 eV for the tetragonal 5$d^1$ DP $Ba_2CaReO_6$ [24], and 0.24 and 0.29 eV for the monoclinic DPs $Sr_2MgOsO_6$ and $Ca_2MgOsO_6$ with $Os^{6+}$ [31], respectively. These small band gaps, despite the large separations between Re or Os ions of ~5.7 Å may be due to the spatially extended $5d$ orbitals.

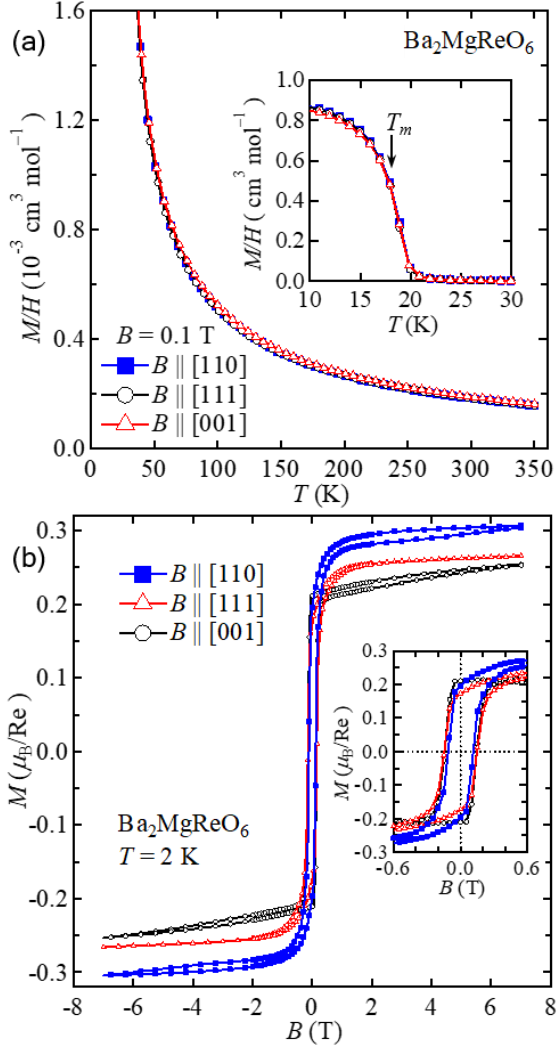

Fig. 3. (Color online) (a) Temperature dependences of the magnetic susceptibility of a $Ba_2MgReO_6$ crystal in magnetic fields of 0.1 T applied along the [110], [111], and [100] directions. The inset shows the temperature dependences of magnetic susceptibility between 10 and 40 K, in which a sharp rise due to a magnetic transition accompanied by a ferromagnetic moment is observed at $T_m$. (b) Field dependences of magnetization of $Ba_2MgReO_6$ at 2 K. The inset panel enlarges an area around the origin, showing hysteresis loops with coercive fields of ~0.15 T.

*2 Magnetization*

To unveil the magnetic properties and their anisotropy of $Ba_2MgReO_6$, magnetic susceptibility measurements were performed on one single crystal sample by applying magnetic fields along the three high-symmetry directions of [100], [110], and [111]. Figure 3(a) shows the temperature dependences of magnetic susceptibility measured using a crystal in a magnetic field of 0.1 T. All the magnetic susceptibilities overlap each other, indicating a weak anisotropy. They exhibit steep increases below ~18 K, which corresponds to the sharp peak at zero field in heat capacity (Fig. 5). Thus, a long-range magnetic order sets in at $T_m$ = 18 K, which is consistent with the previous results using polycrystalline samples [20,30].

The isothermal magnetization curves at 2 K in Fig. 3(b) show ferromagnetic behavior having hystereses with small coercive fields of ~0.15 T. Since the ferromagnetic response appears below $T_m$, which causes the steep increase in magnetic susceptibility, they are associated with the magnetic order at $T_m$. Thus, a ferromagnetic or a canted antiferromagnetic order should occur. The magnitudes of magnetization at 2 K in an applied field of 7 T are 0.254, 0.265, and 0.307 in the unit of $\mu_B$ for magnetic fields along the [100], [111], and [110] directions, respectively, which are close to $0.3\mu_B$ reported for a powder sample at 2 K in an applied field of 5 T [30]. These values seem to be significantly smaller than the saturation moment of the $Re^{6+}$ ion. Moreover, note that there is a significant difference among their values, indicating of small magnetic anisotropy.

There are further notable features in the magnetization curves of Fig. 3(b). The magnetization for [111] tends to saturate at high fields above ~1 T, as expected for a simple ferromagnet, while those for [001] and [110] continuously increase with increasing magnetic field above ~1 T up to 7 T. Moreover, note that there is another hysteresis opening at the high fields only for [001] and [110] in addition to the conventional one at around zero field. These anisotropic and complex isothermal magnetization curves suggest that the spin structure is not simple, which will be discussed later based on the spin model.

The magnetic susceptibility at high temperatures above ~80 K follows the Curie–Weiss (CW) law, as evidenced by the linear behavior in its inverse in Fig. 4. Curie–Weiss fits to the data between 80 and 350 K yield effective magnetic moments of $0.678(1)\mu_B$, $0.689(1)\mu_B$, and $0.673(1)\mu_B$, and Weiss temperatures $\Theta_W$ of −14.6(2), −15.2(3), and −11.2(2) K for the [100], [110], and [111] field directions, respectively. The obtained effective magnetic moments are almost equal to each other and highly reduced from the spin-only value of $1.73\mu_B$ for $S = 1/2$. On the other hand, the magnitudes of the Weiss temperatures are lower than $T_m$ and also the negative sign of them, which indicates dominant antiferromagnetic interactions, seems to be incompatible with the observed ferromagnetic behavior. These facts suggest that competing ferromagnetic and antiferromagnetic interactions are involved, which may cause a complex magnetic order.

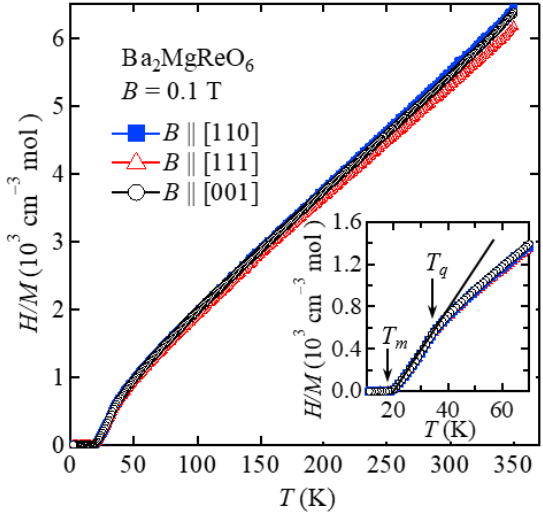

Fig. 4. (Color online) Temperature dependences of the inverse magnetic susceptibilities of a $Ba_2MgReO_6$ crystal in magnetic fields of 0.1 T applied along the [110], [111], and [001] directions. The inset panel expands the low-temperature part, in which a change in the slope of the inverse susceptibilities is observed at around $T_q$. The black line is a guide to eyes.

The magnetic susceptibility obtained in the present study is substantially different from the previous one reported for a polycrystalline sample, which shows a nonlinear behavior with a convex curvature in the inverse susceptibility. An effective moment of $1.51\mu_B$ and a Weiss temperature of $-373$ K were obtained from a CW fitting to the data between 200 and 300 K for the polycrystalline sample [30]. In general, such a nonlinear behavior is observed when a temperature-independent component is relatively large compared with a CW component in the total magnetic susceptibility. Judging from the fact that the present inverse magnetic susceptibility of the single crystal is apparently proportional to temperature in a wide range, the intrinsic temperature-independent term should be negligible. Possibly, the magnetic susceptibility of the polycrystalline sample must have contained a large temperature-independent contribution that may originate from extrinsic source such as certain disorder or impurity phases. We think that the effective moments and Weiss temperatures obtained in the present study should give reliable parameters for $Ba_2MgReO_6$.

The effective magnetic moment significantly reduced from $1.73\mu_B$ evidences the presence of a strong SOC in $Ba_2MgReO_6$. Similarly reduced effective magnetic moments are observed in the related DP compounds with the $d^1$ electron configuration: $0.744\mu_B$ for $Ba_2CaReO_6$ [29], $0.733\mu_B$ for $Ba_2LiOsO_6$ [20], and $0.677\mu_B$ [20] or $0.596$–$0.647\mu_B$ [21] for $Ba_2NaOsO_6$. In general, for the $d^1$ electron configuration, a $J_{eff} = 3/2$ quartet is generated from the combination of the spin momentum $s = 1/2$ and the effective orbital momentum $l_{eff} = -1$ of the $t_{2g}$ manifold; thus, the total magnetic moment $M = 2s + l$ becomes zero [14]. However, when the center ion hybridizes with the ligands, the orbital momentum can be reduced; thus, the cancellation becomes incomplete, resulting in a finite magnetic moment. According to the recent *ab initio* calculation for $Ba_2NaOsO_6$, the reduction of the orbital moment from the ideal value of $l_{eff} = -1$ is evaluated by a scale factor $\gamma$, which is related to the g-factor as the formula $g = 2(1 - \gamma)/3$ [32]. The calculated $\gamma$ of 0.536 for the oxide ligands reproduces $\mu_{eff} \sim 0.6\mu_B$ as observed in the real materials. One expects a similar $\gamma$ value of 0.49 to explain the effective moments of $\sim 0.68\mu_B$ for $Ba_2MgReO_6$. Therefore, the ground state of $Ba_2MgReO_6$ is approximated to a $J_{eff} = 3/2$ ($\Gamma_8$) state [10].

It is noted that the crystal field of cubic symmetry $O_h$ at the Re site is crucial for realizing the $J_{eff} = 3/2$ state in $Ba_2MgReO_6$. In a lower crystal field, the degeneracy of the $t_{2g}$ orbitals is lifted, and the orbital moment is quenched. For example, the oxychloride $Ca_3ReO_5Cl_2$ comprises $Re^{6+}$ ions in a square-pyramidal coordination with five oxide atoms, in which the degeneracy of the $5d$ orbitals is completely lifted by the crystal field [33]. The observed magnetic moment of $Ca_3ReO_5Cl_2$ is $1.585(2)\mu_B$ [34], which is much closer to the spin-only value for $S = 1/2$.

*3 Heat capacity*

Heat capacity measurements on a single crystal reveal a sharp peak at the magnetic transition with $T_m = 18$ K, as shown in Fig. 5, which demonstrates a well-defined phase transition in bulk occurring in a clean crystal. Also, a broad anomaly is observed at a higher temperature centered at $T_q = 33$ K. This must correspond to the broad peak observed in the heat capacity of polycrystalline samples. Thus, it does not originate from an impurity but from an intrinsic phase transition in $Ba_2MgReO_6$.

In order to extract the electronic heat capacity, a lattice contribution is roughly estimated by fitting the high-temperature data to the sum of two Debye functions with different Debye temperatures of 299 and 796 K, which nicely reproduces the data above 80 K (the inset of Fig. 4). By subtracting the lattice contribution, the presence of the broad peak at $T_q$ becomes evident in the curve of the electronic contribution $C_e/T$. The electronic entropy deduced from $C_e/T$ saturates at $S_e = 11.32$ J K$^{-1}$ mol$^{-1}$ above 80 K. This value is much larger than $R\ln2 = 5.76$ J K$^{-1}$ mol$^{-1}$ expected for a doublet of $S = 1/2$ and rather close to $R\ln4 = 11.5$ J K$^{-1}$ mol$^{-1}$ expected for a quartet, which supports the $J_{eff} = 3/2$ quartet realized in $Ba_2MgReO_6$. There is also a contribution from an excited state, a $J_{eff} = 1/2$ doublet, separated by $3/2\lambda$ above the $J_{eff} = 3/2$ quartet, where $\lambda$ is the strength of the SOC. [14] However, because $\lambda$ may be as large as 0.4 eV (several thousands of K) in 5$d$ compounds [12,35,36], the $J_{eff} = 3/2$ quartet is a good description for the ground state in the present temperature range. A smaller electronic entropy of 9.68 J K$^{-1}$ mol$^{-1}$ below 45 K was reported for polycrystalline Ba$_2$MgReO$_6$ [30]. In contrast, the $R$ln4 electronic entropy has not been observed for Ba$_2$NaOsO$_6$: only 4.6 J K$^{-1}$ mol$^{-1}$ was estimated below the magnetic transition temperature [21]. This may be due to the half of the electronic entropy is released at higher temperatures below the quadrupolar transition at 320 K.

The entropies released below $T_m$ and $T_q$ are 2.72 and 6.85 J K$^{-1}$ mol$^{-1}$, which correspond to 47% and 119% of $R$ln2, respectively. Thus, half of the total entropy is approximately involved in the broad peak at $T_q$ and the other half in the sharp peak at $T_m$. This means that the $J_{\text{eff}} = 3/2$ quartet in the paramagnetic state splits into two sets of Kramer's doublets at $T_q$ owing to a quadrupolar ordering and one of the Kramer's doublet at low energy splits at the magnetic transition $T_m$.

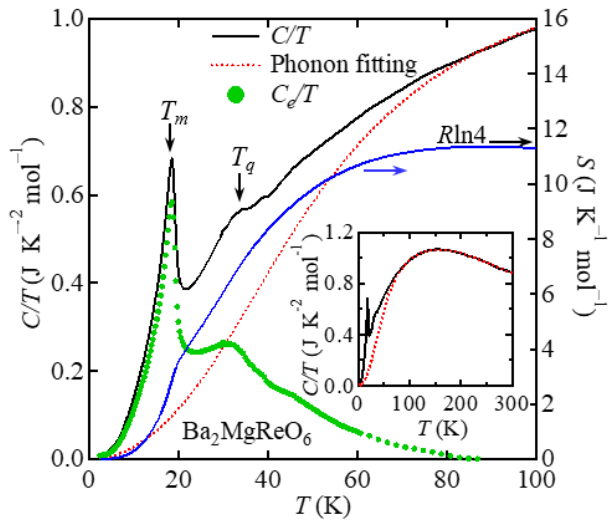

Fig. 5. (Color online) Temperature dependences of heat capacity divided by temperature ($C/T$, black line, left scale), lattice contribution (red line, left scale), and electronic contribution ($C_e/T$, green line, left scale) of a Ba$_2$MgReO$_6$ crystal. The blue line (right scale) represents the electronic entropy ($S$) deduced from the electronic component in $C/T$. The lattice contribution is estimated by fitting the $C/T$ at high temperatures above 80 K to the sum of two Debye functions, as shown in the inset.

## IV. DISCUSSION

### 1 Phase diagram

Chen, Pereira, and Balents have investigated the ground states of the DP with the $d^1$ electron configuration in the presence of strong SOC in a mean field approximation [10,18]. Assuming three kinds of interactions between transition metal ions, that are, an antiferromagnetic exchange interaction $J$, a ferromagnetic interaction through oxygen ions $J'$, and an electrostatic quadrupole interaction $V$, they obtain three ground states; an antiferromagnetic phase with dominant octupolar order in the small $J'$ and small $V$ regime, a CAF phase with the net moment along the [110] direction (CAF [110]) in the large $J'$ and large $V$ regime, and another CAF phase with the weak ferromagnetic moment oriented along the [100] direction (CAF [100]) in the narrow intermediate range between them. Figure 6 illustrates a phase diagram for $J' \sim 0.2J$ as a function of $V/J$. It is suggested that the CAF phases are stabilized by the complex, anisotropic interactions between quadrupoles located at the Re site in the quadrupole order at higher temperatures. Therefore, they appear below the quadrupole phase in the phase diagram of Fig. 6, and the two transitions occur successively with decreasing temperature.

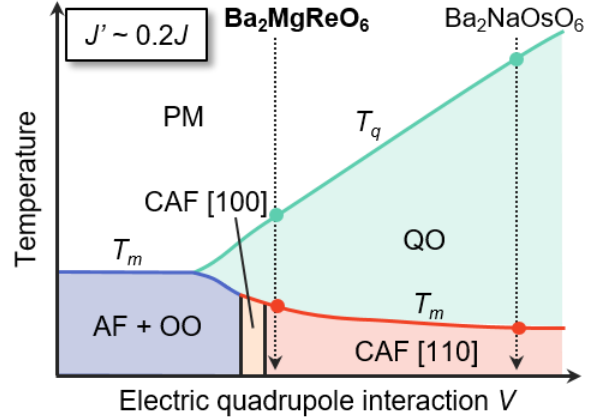

Fig. 6. (Color online) Schematic phase diagram for the double perovskite with the $d^1$ electron configuration as functions of temperature and the electric quadrupole interaction $V$ in the case of $J' \sim 0.2 J$, which is drawn based on ref [10]. For sufficiently large $V$, canted antiferromagnetic (CAF) phases with ferromagnetic moments along the [100] and [110] directions exist below a quadrupolar order (QO) phase. PM and (AF + OO) represent paramagnetic phase and antiferromagnetic order with dominant octupolar order, respectively.

We have shown that Ba$_2$MgReO$_6$ takes a $J_{\text{eff}} = 3/2$ state and exhibits two transitions: possible quadrupolar transition at $T_q = 33$ K and a CAF [110] order below $T_m = 18$ K; the detail and characteristics of them will be addressed in the following sections. Here we roughly estimate where the compound is located in the phase diagram. The negative Weiss temperature suggests that the antiferromagnetic interaction $J$ is larger than the ferromagnetic interaction $J'$. The fact that the $T_q$ is approximately twice as large as the $T_m$ suggests that the electric quadrupole interaction $V$ is also relatively small compared to $J$. The previous study using powder samples estimated the ratio between these parameters as $J : J' : V = 1 : 0.19 : 0.24$ [30]. In the phase diagram of Fig. 6 for $J' \sim 0.2J$, Ba$_2$MgReO$_6$ is located near the edge of the CAF [110] phase at the small $V$ side.

In contrast, Ba$_2$NaOsO$_6$ with $T_q$ = 320 K and $T_m$ = 6.3 K may be located near the large $V$ edge of the phase diagram. Although both compounds have similarly-reduced effective magnetic moments of approximately 0.7$\mu_B$ and exhibit the CAF [110] order with relatively small ferromagnetic moments around 0.3$\mu_B$, the ratios between $T_q$ and $T_m$ seem quite different. This means that the $V$ value is much larger for Ba$_2$NaOsO$_6$ than Ba$_2$MgReO$_6$, the reason of which is not clear. An alternative interpretation could be that the actual quadrupolar order occurs at a lower temperature, not at 320 K but at ~12 K where the NMR experiments observed a structural distortion [19]; the $T_q$ / $T_m$ ratio could be comparable for the two compounds.

### *2 Quadrupolar order*

The complete experimental evidence for the quadrupolar order in Ba$_2$MgReO$_6$ has not yet been attainted. Our observations are a peak in heat capacity and an anomaly in magnetic susceptibility at $T_q$. For the former, it is noted that the entropy estimated below $T_q$ is close to (1/2)$R$ln4, suggesting that the order lifts the $J_{eff}$ = 3/2 quartet into two doublets. The reason why the peak is broad, compared with that at $T_m$, is not known. A similar double peak structure in heat capacity is observed in the $4f^1$ electron compound CeB$_6$, which has a $\Gamma_8$ quartet [37]. This cubic $\Gamma_8$ quartet is identical to the $J_{eff}$ = 3/2 quartet ground state [10] in Ba$_2$MgReO$_6$ in terms of symmetry. Therefore, they have the same multipolar degrees of freedom. In CeB$_6$, an antiferroic quadrupolar order occurs at $T_q$ = 3.3 K, followed by a magnetic transition at $T_N$ = 2.3 K [38–44]. The corresponding peaks in heat capacity are broad at $T_q$, while sharp at $T_N$ [45], just as observed in the present compound. The broad quadrupolar transition in CeB$_6$ is considered to originate from the short-range nature of the quadrupolar interaction, compared with long-range RKKY magnetic interactions, which can be easily disturbed by disorders originated from crystalline defects. For Ba$_2$MgReO$_6$, however, both the quadrupolar and magnetic interactions must be short-ranged in the insulating state. It is likely that the former is more influenced by disorder than the latter owing to certain reason.

Concerning the anomaly in magnetic susceptibility at $T_q$, the inverse magnetic susceptibility in the inset of Fig. 4 shows a distinct kink at $T_q$, below which another linear region appears down to $T_m$. This indicates that magnetic interactions are substantially changed by the quadrupolar order. A CW fitting to the data between 21 and 35 K yields $\mu_{eff}$ = 0.459(3)$\mu_B$, 0.472(4)$\mu_B$, and 0.478(5)$\mu_B$ and $\Theta_W$ = 20.2(1), 20.0(1), and 19.7(2) K for the field directions along [100], [110], and [111], respectively. Compared to the high-temperature CW region with $\mu_{eff}$ ~ 0.68$\mu_B$ and $\Theta_W$ ~ −15 K, the effective moment is reduced to ~70%, and the sign of the Weiss temperature is inverted to positive, indicating that the net interaction dramatically changes from antiferromagnetic to ferromagnetic upon cooling across $T_q$.

The appearance of two CW regimes in magnetic susceptibility has been theoretically predicted and is taken as evidence for a quadrupolar order [10]. In the quadrupolar ordered phase, the $J_{eff}$ = 3/2 quartet splits into two Kramer's doublets with $J_{eff}$ = ±1/2 and ±3/2. Since the lower-energy doublet in the quadrupolar ordered phase has an effective moment different from that of the $J_{eff}$ = 3/2 quartet in the high-temperature paramagnetic phase, the slope of the inverse magnetic susceptibility changes at $T_q$. Recent theory predicts a reduction by a factor of $\sqrt{3}/\sqrt{5}$ (~77%) [46], which is in good agreement with our experimental observation for Ba$_2$MgReO$_6$. On the other hand, the origin of the sign change in the Weiss temperature is difficult to understand. Probably, the balance between competing ferromagnetic and antiferromagnetic interactions is slightly modified at the transition. The microscopic origin should be addressed in the future study.

There remains an open question on the presence of a structural change at the $T_q$ that is induced by a coupling between the quadrupolar order and lattice. It was not observed in the previous powder neutron diffraction experiments [30]. This is possibly because the quadrupolar order is antiferroic and, thus, the lattice distortion may be undetectably small. Another possibility is that disorder or surface strain exerted on the powder sample has smeared out the signature of structural distortion accompanied by the quadrupolar order. In the case of Ba$_2$NaOsO$_6$, the details about the structural distortion has not yet been reported. We plan to carry out high-resolution XRD experiments in order to detect the distortion and to unveil the nature of the quadrupolar order. Moreover, the observation of superlattice reflections by resonant X-ray scattering experiments and the detection of phonon softening by ultrasonic experiments are in progress.

### *3 Magnetic order*

The observed magnetic properties of Ba$_2$MgReO$_6$ are reasonably explained by the CAF spin model proposed for Ba$_2$NaOsO$_6$ [18]. In fact, the magnitudes of the effective and saturation moments are similar between the two compounds: ~0.68$\mu_B$ (~0.47$\mu_B$ below $T_q$) and ~0.3$\mu_B$ for Ba$_2$MgReO$_6$ and ~0.6$\mu_B$ and ~0.25$\mu_B$ for Ba$_2$NaOsO$_6$ [21], respectively. As depicted in Fig. 1, the model consists of the α and β planes with ferromagnetically aligned spins lying in each plane alternately stacked along the [001] direction. The directions of the spins in the α and β planes rotate by angles of +$\phi$ and −$\phi$ with respect to the [110] direction; thus, their directions are symmetric about the (110) mirror plane. This spin canting gives an uncompensated moment of magnitude depending on the angle along the [110] direction. The $\phi$ of Ba$_2$NaOsO$_6$ is estimated to be 67° from the NMR measurements [19].

The isothermal magnetization curves at 2 K in Fig. 3(b) are now discussed based on the CAF spin model, which may carry important information on the quadrupolar and magnetic orders. The magnetization saturates above ~1 T when the field is applied along the [111] direction, while keeps increasing particularly for [001]. In addition, hystereses appear even above 2 T for [001] and [110]. The large anisotropy in the magnitude of magnetization at 7 T is also noted. All these features suggest that a highly anisotropic magnetic order occurs, which must be induced by the preceding quadrupolar order [10]; the hystereses must come from variations in the distribution of magnetic domains.

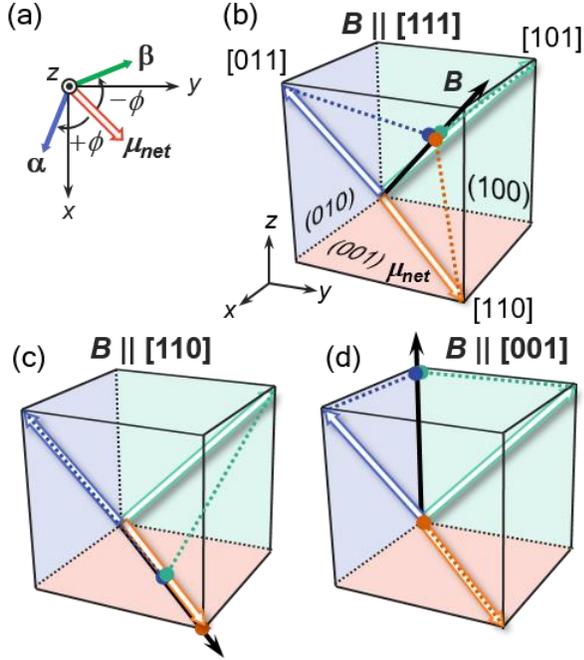

Fig. 7. (Color online) (a) Geometric relationship between spins in the α (blue arrow) and β (green arrow) sublattices and the uncompensated net moment $\mu_{net}$ (red open arrow) along the [110] direction in the tetragonal domain Z, where $\phi$ is the spin canting angle from the [110] direction. (b-d) Schematic representations of the relation between the applied magnetic field $B$ (black arrow) and net ferromagnetic moments $\mu_{net}$ (blue, orange, and green open arrows). For each of the three tetragonal domains generated in the quadrupolar order, one of four magnetic domains that is stabilized by the Zeeman effect is shown: the magnetic domains with $\mu_{net}$ along [110], [011], and [101] for tetragonal domains Z, Y, and X, respectively. Projections of the net moments of the magnetic domains to the field direction are depicted by dotted lines; for example, in (b) with $B \parallel [111]$, each domain contributes equally, giving $\mu_{net} \cos(54.7°)$ along the field direction.

The quadruplar order should reduce the crystal symmetry from cubic to tetragonal. In the tetragonal phase, three types of domains having the $c$ direction along [100], [010], and [001] of the cubic structure are formed (the X, Y, and Z domains, respectively). When the coplanar CAF order takes place, four magnetic domains with the net moments along <110> are generated for each tetragonal domain; for example, those with the net moments along [110], [–1–10], [1–10] and [–110] for the Z domain. Note that the magnetic moments and thus the net moments lie in the (001) plane in the CAF order, which is due to the ordering of the quadrupole moments [18]. Depending on the direction of the applied magnetic field, some of the magnetic domains are stabilized by the Zeeman energy associated with the net moment.

Let us assume that the distribution of the tetragonal domains is not influenced by magnetic field: three of them are always evenly contained. Figure 7 illustrates the geometrical relation between the net moments of magnetic domains and the applied magnetic field. The simplest case is when the magnetic field is applied along the [111] direction: for $B \parallel [111]$, all the three tetragonal domains are equivalent, and a single, equivalent magnetic domain is selected for each tetragonal domain, as depicted in Fig. 7(b); for example, the [110] magnetic domain for the Z tetragonal domain. The resulting moment projected along $B$ from each magnetic domain is $\mu_{net} \cos q$ ($q = 54.7°$), where $\mu_{net}$ is the uncompensated net moment per Re atom. The saturation magnetization $M_s = 0.265\mu_B$ at 7 T for $B \parallel [111]$ yields $\mu_{net} = 0.459\mu_B$. On the other hand, for $B \parallel [110]$ in Fig. 7(c), only the [110] magnetic domain is selected for the Z domain, while the [011] and [01–1] ([101] and [10–1]) magnetic domains for the X (Y) domains. Thus expected net moment is $(1 + 1/2 + 1/2)/3 = 2/3$ in the unit of $\mu_{net}$, which is to be equal to $M_s = 0.307\mu_B$ at 7 T for $B \parallel [110]$ and $\mu_{net} = 0.461\mu_B$ is attained. Therefore, the $M_s$ values for $B \parallel [111]$ and [110] are consistently explained with $\mu_{net} = 0.46\mu_B$, assuming invariant tetragonal domains.

A similar consideration for $B \parallel [001]$, shown in Fig. 7(d), suggests that the [011] and [0–11] ([101] and [–101]) magnetic domains for the X (Y) domains, while all the four magnetic domains coexist for the Z domain (the net moment is always perpendicular to $B$). Provided $\mu_{net} = 0.46\mu_B$ from the above discussion, $M_s$ is calculated to be $0.217\mu_B$, which is close to the magnetization at 1 T for $B \parallel [001]$. Note that the magnetization increases gradually with further increasing field and exhibits a hysteresis. This is probably because of misalignment of the crystal that eventually selects one or two magnetic domains with the net moment along the field direction out of four domains for the Z domain; since the effective field on the domains is small, this additional magnetization increases gradually with field. The hysteresis originates from a variation in the domain distribution between field sweeps rising up and down. Therefore, all the observed features in the magnetization are completely understood in terms of the CAF spin structure stabilized by the quadrupolar order.

The magnitude of the magnetic moment in $Ba_2MgReO_6$ is not determined; neutron diffraction or NMR measurements are required. If one assumes $0.6\mu_B$ as in $Ba_2NaOsO_6$, the canting angle could be 40º, smaller than 67º for $Ba_2NaOsO_6$. The detail of the CAF order as well as the quadrupolar order in $Ba_2MgReO_6$ will be examined in future work.

## V. SUMMARY

We focus on the DP compound $Ba_2MgReO_6$ with the $5d^1$ electron configuration as a promising candidate hosting novel electronic phases derived from the strong SOC of the $5d$ electrons. Thermodynamic measurements using high-quality single crystals reveal that $Ba_2MgReO_6$ is a spin–orbit-coupled Mott insulator with a CAF order accompanied by an uncompensated ferromagnetic moment along the [110] direction below $T_m = 18$ K and with a quadrupolar order below $T_q = 33$ K. Markedly, the inverse magnetic susceptibility changes its slope at $T_q$, indicating that magnetic interactions are significantly modified in the quadrupolar order. These results are interpreted assuming a $J_{eff} = 3/2$ quartet, which is evidenced by the highly reduced effective magnetic moment of $0.68\mu_B$ and the total magnetic entropy of $R\ln 4$. Our findings, together with similar results for $Ba_2NaOsO_6$, are perfectly consistent with the theoretical predictions for the $5d^1$ DPs.

## ACKNOWLEDGEMENTS

The authors are grateful to L. Balents and H. Ishizuka for insightful discussion on the theoretical aspect, and T. Manjo and H. Sawa for helpful discussion on the structural property. This work was partly supported by Japan Society for the Promotion of Science (JSPS) KAKENHI Grant Number JP18K13491, JP18H04308 (J-Physics) and by Core-to-Core Program (A) Advanced Research Networks.